\documentclass[preprint,showpacs,preprintnumbers,amsmath,amssymb,nofootinbib,aps]{revtex4}
\pdfoutput=1
\usepackage{epsfig}

\usepackage{graphicx}
\usepackage{dcolumn}
\usepackage{bm}
\usepackage{amsmath}
\usepackage{latexsym}
\usepackage{color}
\usepackage{hyperref}
\usepackage{mathrsfs}
\usepackage{multirow}

\newcommand{\be}{\begin{equation}}
\newcommand{\ee}{\end{equation}}
\newcommand{\bea}{\begin{eqnarray}}
\newcommand{\eea}{\end{eqnarray}}

\begin{document}

\title{Can dark matter--dark energy interaction alleviate the Cosmic Coincidence Problem?}

\author{J. F. Jesus$^{1,2}$}\email{jf.jesus@unesp.br}
\author{A. A. Escobal$^{2}$} \email{anderson.aescobal@gmail.com}
\author{D. Benndorf$^{2}$}\email{douglas.benndorf@unesp.br}
\author{S. H. Pereira$^{2}$}\email{s.pereira@unesp.br}

\affiliation{$^1$Campus Experimental de Itapeva \\Universidade Estadual Paulista (UNESP),\\  R. Geraldo Alckmin 519, 18409-010, Itapeva, SP, Brazil,\\
\\$^2$Departamento de F\'isica\\ Faculdade de Engenharia de Guaratinguet\'a, \\Universidade Estadual Paulista (UNESP),   Av. Dr. Ariberto Pereira da Cunha 333, 12516-410, Guaratinguet\'a, SP, Brazil}


\begin{abstract}
In this paper we study a model of interacting dark energy -- dark matter where the ratio between these components is {not constant, changing} from early to late times in such a way that the model can solve or alleviate the cosmic coincidence problem (CP). The interaction arises from an assumed relation {of the form}  $\rho_x\propto\rho_d^\alpha$, {where $\rho_x$ and $\rho_d$ are the energy densities of dark energy and dark matter components, respectively,} and $\alpha$ is a free parameter. For a dark energy equation of state parameter $w=-1$ {we found that}, if $\alpha=0$, the standard $\Lambda$CDM model is recovered, where the coincidence problem is unsolved. For $0<\alpha<1$, the CP would be alleviated and for $\alpha\sim1$, the CP would be solved. The dark energy component is analyzed {with both} $w=-1$ and $w\neq -1$. {Using Supernovae type Ia and Hubble parameter data constraints},  in the case $w=-1$ {we find} $\alpha=0.109^{+0.062}_{-0.072}$ at 68\% C.L., and the CP is alleviated. For $w\neq-1$, a degeneracy arises on the $w$ -- $\alpha$ plane. {In order to break such degeneracy we add cosmic microwave background distance priors and baryonic acoustic oscillations data to the constraints, yielding $\alpha=-0.075\pm0.046$ at 68\% C.L.. In this case we find that the CP is not alleviated even for 2$\sigma$  interval for $\alpha$. Furthermore, this last model is discarded against {flat} $\Lambda$CDM according to BIC analysis.}

\end{abstract}

\maketitle



\section{Introduction}

The standard model of cosmology, known as $\Lambda$CDM model, correctly describes the type Ia Supernovae (SNe Ia) observations,
 which indicate a recent accelerated expansion of the universe. It also explains quite well the formation of large-scale structures and the cosmic abundance of different types of matter and energy. However, although being a model extremely predictive and observationally robust, $\Lambda$CDM suffers from profound theoretical difficulties, such as not giving a reasonable explanation for the nature of cold dark matter (CDM) and the very discrepant value between the observed and predicted for cosmological constant $\Lambda$. {For an interesting and recent review on $\Lambda$CDM problems we refer the reader to \cite{Bull2016}. } Another very recent problem concerning the standard model is the so called Hubble tension, a statistical significant disagreement between predictions of $H_0$ by early time probes against a number of late time determinations of $H_0$ from local measurements of distances and redshifts (see \cite{valentino2021} for a review).

{ There are several alternative models to $\Lambda$CDM. An overall class of theories replace the constant $\Lambda$ term by a dynamical dark energy (DE) component (see \cite{GongBo} and \cite{marttens2020} for a recent discussion).} Other consider the possibility of a coupling between dark matter (DM) and dark energy \cite{Subinoy2005,SF, Cabral2009, Majerotto2009,Valiviita2010,Chimento2010,Cai2010,Sun2012,Pourtsidou2013,Salvatelli2014,Li2014,Skordis2015,Jimenez2016,Shafieloo2018,Valent2020}. {This specific class of theories have the advantage to explain why the present values of dark energy and dark matter densities are of the same order of magnitude}, which indicate we are living in a very special moment of the cosmic history. This is known as the cosmic Coincidence Problem (CP) \cite{Huey2004,Velten2014}. In \cite{marttens2020} it was shown that a dynamical DE model and the interacting
DM-DE approaches become indistinguishable both at the background and linear perturbation level. Thermodynamic properties of interaction models were studied in \cite{SF,Sun2012}, showing that the presence of a non null
chemical potential for at least one of the fluids allows
the decay from the DM fluid into DE, with no violation
of the second law of thermodynamics and also being favored by cosmological data in some cases. In \cite{Salvatelli2014} it was shown that a general late-time interaction between cold dark matter and vacuum energy is also in agreement to
current cosmological data sets, occurring at about $z=0.9$. A new class of metastable dark energy phenomenological models in which the DE decay rate does not depend on external parameters was proposed in \cite{Shafieloo2018}. A scenario in which DM particles interact via a force mediated by
a scalar field was proposed very recently in \cite{Valent2020}, where the scalar field drives cosmic acceleration.

Most of the interacting DM-DE models are described by the Friedmann equations and the conservation {equations}:
\begin{equation}
    \dot{\rho}_d + 3H(\rho_d + p_d)=Q\,,\label{rhoDM}
\end{equation}
\begin{equation}
\dot{\rho}_x + 3H(\rho_x + p_x)=-Q\,,\label{rhoDE}
\end{equation}
where $\rho_d$ and $p_d$ are DM energy density and pressure, and $\rho_x$ and $p_x$ are DE energy density and pressure. For a dust like DM fluid we have $p_d=0$ and for a vacuum like DE fluid we have $p_x =-\rho_x$, for instance.  $Q$ denotes the phenomenological interaction term. For $Q>0$ we have DE decaying into DM while for $Q<0$ we have DM decaying into DE. A thermodynamic analysis \cite{SF} showed that both decaying route are possible. 

Several interaction terms $Q$ have been studied in the literature {\cite{SF,Cabral2009,Majerotto2009,Valiviita2010,Chimento2010,Cai2010,Sun2012,Pourtsidou2013,Salvatelli2014,Li2014,Skordis2015,Jimenez2016,Shafieloo2018,Valent2020}}. Most of them are $Q \propto \rho_x$ or $Q \propto \rho_d$. Models with $Q \propto H \rho_d$ have been largely explored in literature, however it is plagued with instabilities at the evolution of density perturbations {\cite{Valiviita2008,Jian2009}}. Models with $Q \propto H \rho_x$ are free of such instabilities. In particular, the model with $Q = 3\varepsilon H \rho_x$ has been studied in \cite{SF} and the $\varepsilon$ parameter has been {constrained by} SNe Ia observational data as $\varepsilon = -0.026^{+0.021}_{-0.027}$ at $1\sigma$ C.L. This shows that a negative value of $Q$ {is favoured by observations at $1\sigma$ interval, with the $\varepsilon = 0$ limit only slightly away from the best-fit value.}

Following a different approach, Cai and Su \cite{Cai2010} have found that there must be a change of the sign of $Q$ at about $z=0.5$ in a model independent of the specific interaction term. This raises a remarkable challenge to the interacting models, since the usual phenomenological forms of interaction do
not change their signs during the cosmological evolution. A model that encompass such very peculiar characteristic is presented in \cite{Sun2012}.

Very recently a model with $Q=\Gamma \rho_x$ was presented by Shafieloo et al. \cite{Shafieloo2018}, where $\Gamma$ is a {constant with the dimensions of inverse
time, related to decaying of DE}. With a vacuum like equation of state, $p_x=-\rho_x$, Eq. (\ref{rhoDE}) has a solution of the form $\rho_x(t)=\rho_x(t_0)\exp[-\Gamma(t-t_0)]$, exactly like a radioactive decay of matter and $\Gamma$ is
related to the ‘half-life’ of DE. Such model is known as a metastable dark energy with radioactive-like decay. {A more recent analysis in light of Planck 2018 data was done in \cite{Yang2020}, and a dynamical system analysis was done in \citep{Szy2020}. A very interesting interacting closed model has been explored recently with full CMB data, which offer a very compelling solution
to the Hubble constant tension \cite{Yang2021,Valentino2021}}.

In the present paper we analyse a model of interacting DM-DE fluids in which we aim to test if an interaction between DM and DE may solve or alleviate the CP. In $\Lambda$CDM model, the cosmological constant density has always the same value, namely $\rho_\Lambda = \rho_{\Lambda 0}$, while the energy density of pressureless matter satisfies $\rho_d=\rho_{d0}(a/a_0)^{-3}$, where the subscript "0" stands for present day values. The CP says that $\rho_{\Lambda 0} \sim \rho_{d0}$ today. Since that the ratio between the scale factor today to that one at last scattering surface is $a_0/a = 10^3$ or $z\sim 1000$, we have a ratio $r=\frac{\rho_d}{\rho_{\Lambda}}\sim10^{9}$ at that epoch. {Thus, an interaction between DM-DE could maintain the ratio constant along the evolution, $r\sim 1$, solving the CP, since it would not be specific to the present moment.} If the ratio is such that $r<<10^9$ at $z\sim 1000$ we say that CP is alleviated.

\begin{figure}
    \centering
    \includegraphics[width=.49\textwidth]{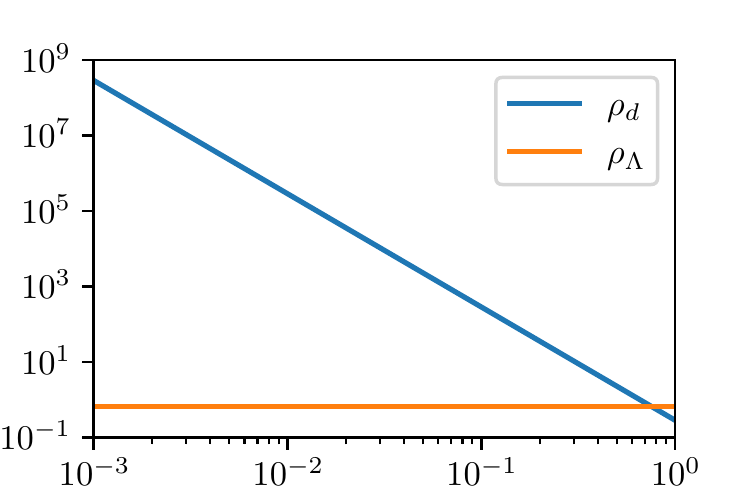}
    \includegraphics[width=.49\textwidth]{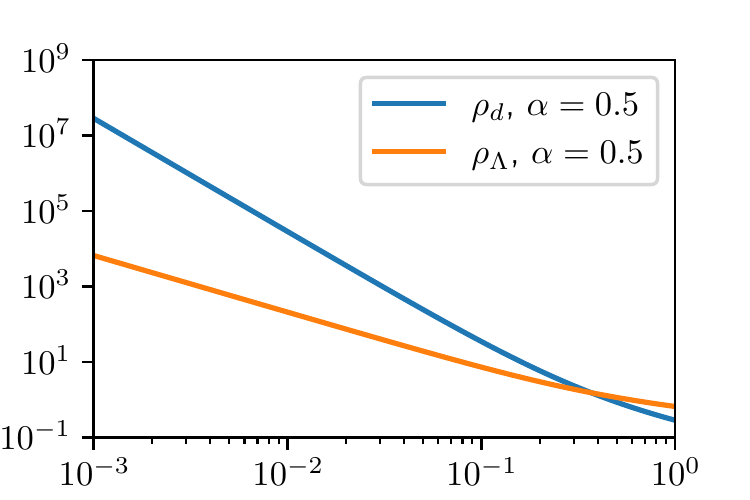}
\caption{Dark sector densities evolution for $\Lambda$CDM (left) and $\Lambda(t)$CDM (right) models. {The densities are normalized by current critical density, $\rho_{c0}$.}}
    \label{fig:evo}
\end{figure}

{The standard $\Lambda$CDM model is characterized by a constant equation of state parameter $w=\frac{p_\Lambda}{\rho_\Lambda}=-1$, where both $\rho_{\Lambda}$ and $p_{\Lambda}$ are constant.} The dark matter and $\Lambda$ densities evolutions can be seen in logarithmic scale in Fig. \ref{fig:evo} (left). The logarithmic densities in this case are given by
\begin{align} 
    \ln\rho_\Lambda=\ln\rho_{\Lambda0},\quad\ln\rho_d=-3\ln a+\ln\rho_{d0}.
\end{align}
{For models where the equation of state parameter is also -1 but with both $\rho_\Lambda$ and $p_\Lambda$ time varying, the energy density associated to $\Lambda$ changes with time, and this defines a $\Lambda(t)$CDM model, with a typical behaviour for the evolution of densities given in Fig. \ref{fig:evo} (right).
If the energy density of $\Lambda(t)$ approximates the dark matter energy density in Fig. \ref{fig:evo}, we solve or at least alleviate the CP. For this,} let us try to write:
\begin{align}
    \ln\rho_\Lambda=k_1\ln a + \ln\rho_{\Lambda0}, \quad\ln\rho_d=k_2\ln a+\ln\rho_{d0}\,.
\end{align}
Rearranging {the equations gives}:
\begin{align}
    \frac{\rho_\Lambda}{\rho_{\Lambda0}}=\left(\frac{\rho_d}{\rho_{d0}}\right)^{k_1/k_2}\,,
\end{align}
which corresponds to:
\be
\rho_x=\beta\rho_d^\alpha
\label{rhoxd}
\ee
for a general DE component.

Thus, in order to test if an interaction model can solve or alleviate the CP, we assume the relation \eqref{rhoxd} for the dark sector. With this assumption, we have that if the data indicate $\alpha\sim0$, there is no interaction and $\Lambda$CDM model is recovered. If {$\alpha = 1$}, the ratio $r$ is exactly constant and the CP is solved. For $0<\alpha<1$, we may conclude that the DM-DE interaction may alleviate the CP. In Fig. \ref{fig:evo} (right), we see how $\alpha=0.5$ alleviates the CP. Below we derive which interaction term $Q$ results in the relation \eqref{rhoxd}.


In Section II the main equations are presented, in Section III the model is analysed and in Section IV we have the conclusions.

\section{DE-DM interacting model}

The Friedmann equation in a flat background for the present model is:
\begin{equation}
    H^2=\frac{\kappa^2}{3}(\rho_b+\rho_r+\rho_d + \rho_x)\,,\label{H2}
\end{equation}
where $\kappa^2 = 8\pi G$, and the conservation laws for the baryonic energy density $\rho_b$, radiation energy density $\rho_r$, DM energy density $\rho_d$ and DE energy density $\rho_x$ are
\begin{equation}
    \dot{\rho}_b + 3H\rho_b=0\,,\label{b}
\end{equation}
\begin{equation}
    \dot{\rho}_r + 4H\rho_r=0\,,\label{rad}
\end{equation}
\begin{equation}
    \dot{\rho}_d + 3H\rho_d =Q\,,\label{DMQ}
\end{equation}
\begin{equation}
\dot{\rho}_x + 3H(1+w)\rho_x =-Q\,,\label{DEQ}
\end{equation}
where we assume that {baryons} and dark matter satisfy a dust like equation of state, $p_r=\frac{\rho_r}{3}$ and dark energy satisfies
\begin{equation}
    p_x=w \rho_x\,,\label{omega}
\end{equation}
where $w$ is the DE equation of state (EOS) parameter.

Inspired by the metastable dark energy decay model presented above we chose to work with a phenomenological interaction term of the form
\begin{equation}
    Q=\Gamma(t)\rho_x\,,\label{Qrhox}
\end{equation}
where now $\Gamma(t)$ represents a time dependent decaying rate and the energy densities of DE and DM are related by \eqref{rhoxd}:
\begin{equation}
    \rho_{x}=\beta\rho_{d}^\alpha\equiv  f(\rho_{d})\,,
    \label{rhox}
\end{equation}
with constant $\alpha$ and $\beta$, such that for  $\alpha \neq 1$ the $\beta$ parameter is a {dimensionful} constant, and for $\alpha = 1$ we see that the ratio $\rho_x/\rho_d$ is just a constant, $\beta$, showing that both densities evolves exactly at the same manner. A small deviation of $\alpha$ from unity shows a nontrivial dependence between DM and DE densities.

From (\ref{DEQ}), (\ref{Qrhox}) and (\ref{rhox}) we obtain:
\begin{equation}
    \Gamma(t)=-3H(1+w)-\alpha\frac{\dot{\rho}_{d}}{\rho_{d}}\,,
    \label{Gamma}
\end{equation}
and from (\ref{DMQ}):
\begin{equation}\label{oderhodw}
   \frac{\dot{\rho}_{d}}{\rho_d} \Bigg[\frac{1+\alpha \beta\rho_d^{\alpha-1}}{1+\beta(1+w)\rho_d^{\alpha-1}}\Bigg] = -3H = -3 \frac{\dot{a}}{a}\,.
\end{equation}
The above equation can be integrated to obtain $\rho_d(a)$ in an implicit logarithm equation, which is useful just for the particular case $w=-1$. For the general case, Eq. (\ref{oderhodw}) must be solved numerically.

\subsection{The case \texorpdfstring{$w\neq -1$}{w!=-1}}

In the more general case, $w\neq-1$, Eq. \eqref{oderhodw} can not be solved analytically to obtain $\rho_d(a)$. So, in this case, we choose to solve numerically the differential equations \eqref{DMQ}-\eqref{oderhodw} to obtain $\rho_d$, $\rho_x$ and $H$ as function of the scale factor or redshift.

\begin{figure}[h!]
    \centering
    \includegraphics[width=.49\textwidth]{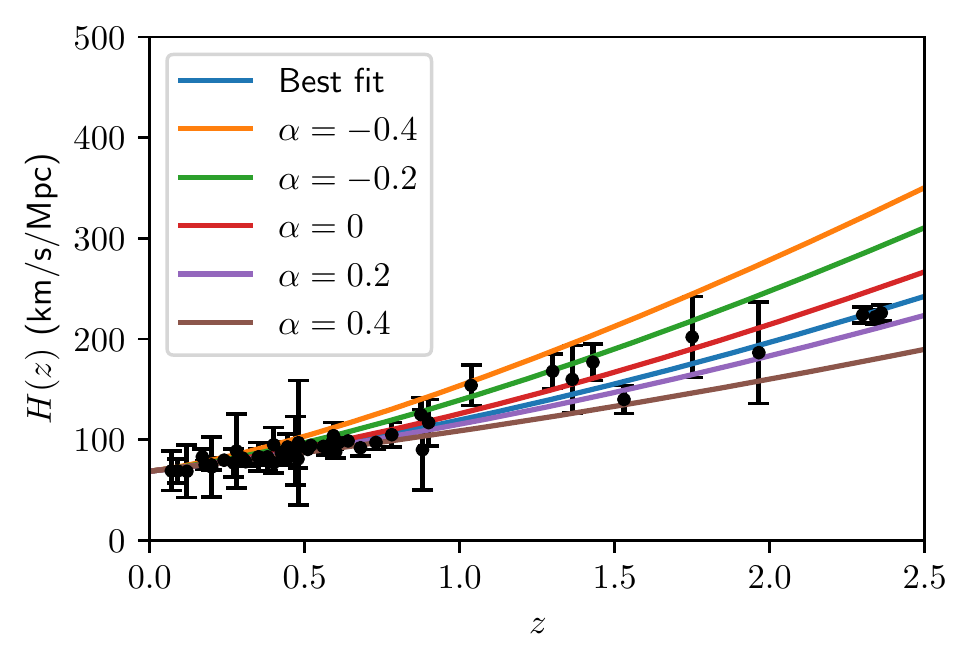}
    \includegraphics[width=.49\textwidth]{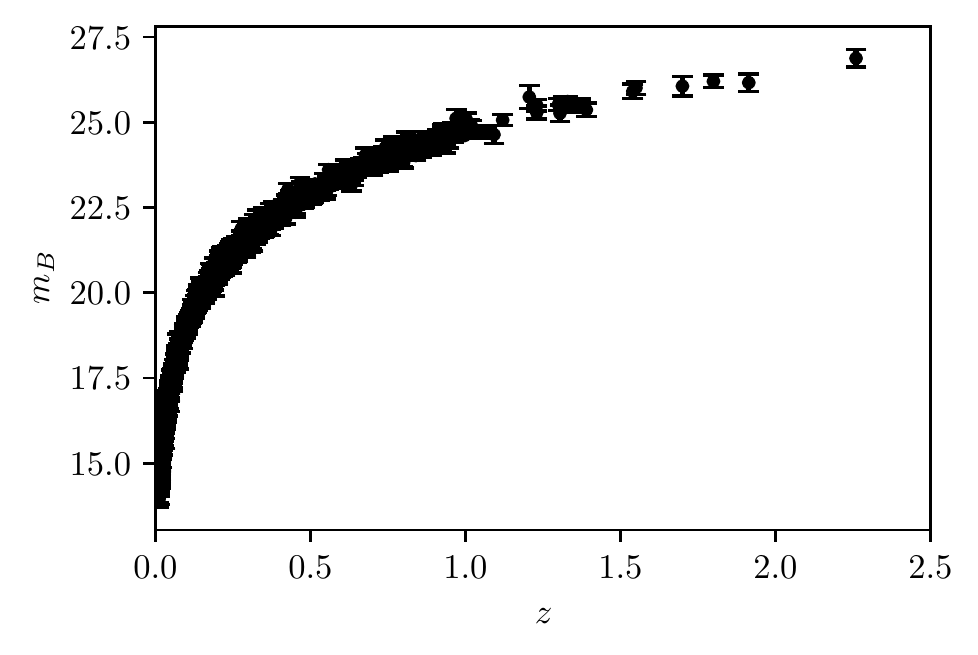}
    \caption{{\bf Left:} $H(z)$ data \cite{MaganaEtAl17} and some curves predicted from the $\Lambda(t)$CDM model for some values of $\alpha$. The best fit corresponds to Table \ref{tab:LtCDM}. $\alpha=0$ corresponds to standard $\Lambda$CDM model. {\bf Right:} Apparent SNe Ia magnitudes $m_B$ from Pantheon compilation \cite{pantheon}.}
    \label{fig:data}
\end{figure}

Aiming to constrain the model with data, we rewrite Eq. \eqref{oderhodw} in terms of the redshift and in terms of dimensionless quantities. So, by using the definitions $r_d\equiv\frac{\rho_d}{\rho_{d0}}$, $r_x\equiv\frac{\rho_x}{\rho_{x0}}$, $\Omega_{d0}\equiv\frac{\rho_{d0}}{\rho_{c0}}$, $\Omega_{x0}\equiv\frac{\rho_{x0}}{\rho_{c0}}$, {where $\rho_{c0}$ is the present critical density}, and the relation $\frac{d}{dt}=-H(1+z)\frac{d}{dz}$, we may rewrite Eq. \eqref{oderhodw} as:
\be\label{drddz}
\frac{dr_d}{dz}=\frac{3r_d}{1+z}\left[\frac{\Omega_{d0}+(1+w)\Omega_{x0}r_d^{\alpha-1}}{\Omega_{d0}+\alpha\Omega_{x0}r_d^{\alpha-1}}\right]
\ee
with initial condition $r_d(z=0)=1$. Relation \eqref{rhox} now is simply:
\be
r_x=r_d^\alpha
\ee
and
\be
E(z)\equiv \frac{H(z)}{H_0}=\left[\Omega_{b0}(1+z)^3+\Omega_{r0}(1+z)^4+\Omega_{d0}r_d(z)+\Omega_{x0}r_x(z)\right]^{1/2}\,\label{Ez}
\ee
can be used to constrain the model with observational data and we have included the baryon and radiation contributions, $\Omega_{b0}\equiv\frac{\rho_{b0}}{\rho_{c0}}$ and $\Omega_{r0}\equiv\frac{\rho_{r0}}{\rho_{c0}}$, respectively. Due to spatial flatness, {for $z=0$ the Eq. (\ref{Ez})} reads $\Omega_{b0}+\Omega_{r0}+\Omega_{d0}+\Omega_{x0}=1$.

\subsection{The case \texorpdfstring{$w = -1$}{w=-1}}

The case $w=-1$ can be solved analytically with the aid of the Lambert $w$ function, as we show in details in Appendix A. The solution for $\rho_d(z)$ is:
\be
\rho_d=\rho_{d0}\left[\frac{W\left(\alpha r_0 e^{\alpha r_0} (1+z)^{3(\alpha-1)}\right)}{\alpha r_0}\right]^{\frac{1}{\alpha-1}}
\ee
In this case,
\be
\rho_x(z)=\rho_{x0}\left[\frac{W\left(\alpha r_0 e^{\alpha r_0} (1+z)^{3(\alpha-1)}\right)}{\alpha r_0}\right]^{\frac{\alpha}{\alpha-1}}
\ee
and
\be
E^2 =\frac{H(z)^2}{H_0^2}=\Omega_{b0}(1+z)^3+\Omega_{r0}(1+z)^4+\frac{\rho_d+\rho_x}{\rho_{c0}}\,
\ee
is used to constrain the model with observational data.


\section{Data}
The cosmological data used to constrain the models are briefly described below.

\subsection{\texorpdfstring{$H(z)$}{H(z)} data}
{The Hubble parameter data, named $H(z)$,} can be {obtained} from many sources, with the main being clustering ({Luminous Red Galaxies (LRGs) and Baryon Acoustic Oscillations (BAO)}) and differential age of objects (cosmic chronometers). Here we use the largest known sample of $H(z)$ data up to date, with 51 measurements \cite{MaganaEtAl17}. This dataset can be seen in Fig. \ref{fig:data} (left). We use these data for the $\Lambda(t)$CDM analysis. {For DM-DE interacting model we include separate BAO data in the analysis. In order to avoid unknown correlations with this BAO data, we exclude $H(z)$ data coming from clustering estimates,} reducing the compilation to 31 cosmic chronometers. This separation can be seen in Table 1 of Ref. \cite{MaganaEtAl17}.

\subsection{SNe Ia}
SNe Ia luminosity distances are good tracers of the late Universe history, thereby yielding {strong} {constraints on} dark energy. Here we use the largest sample up to date, namely the Pantheon sample \cite{pantheon}, which consists of 1048 SNe apparent magnitudes over the redshift range $0.01<z<2.3$. This dataset can be seen in Fig. \ref{fig:data} (right).

\subsection{Cosmic Microwave Background}
The Cosmic Microwave Background (CMB) yields strong {constraints on} cosmological models. Acting basically as a standard ruler at the {last scattering surface}, it consists of a strong constraint because it integrates the Universe history from today ($z=0$) up to $z\sim1000$. Here we work with the so called CMB distance priors, which include quantities that are representative of the full CMB spectrum. We choose to work with the priors on the shift parameter, which is related to the position of the first acoustic peak in the power {spectrum of the CMB anisotropies}, acoustic scale and baryon density, ($R,\ell_A,\Omega_bh^2$), respectively, as described by \cite{ChenEtAl19}. We work with the priors coming from Planck (2018) \cite{Planck18}, as indicated by Table 1 of \cite{ChenEtAl19}. The CMB temperature and radiation density have been fixed from \cite{Fixsen09}.

\subsection{BAO}
Prior to the recombination, baryons and radiation were strongly coupled. Due to this coupling, baryon clustering was suppressed by photon pressure. These oscillations between baryon clustering and photon pressure repulsion resulted in an imprint on the correlation function of galaxies after recombination. The correlation function of galaxies decays with distance, as expect due to gravity being an attractive force. The baryon acoustic oscillations, however, results in an excess at the correlation function at the scale of sound horizon at recombination, $r_s\sim150$ Mpc. The position of this BAO signature imposes strong {constraints on} the matter density. Here we use the BAO signature estimate from various sources, as indicated in Tables II and III of Ref. \cite{CamarenaMarra18}.

\section{Analysis}
{In all analyses here, we have used the free software emcee \cite{ForemanMackey13,GoodWeare} in order to probe the posterior distributions $p\propto\pi\mathcal{L}$, where $\pi$ is the prior and $\mathcal{L}\propto e^{-\chi^2/2}$ is the likelihood. The assumed priors were flat with large intervals on parameters encompassing all the non-negligible region of the likelihoods, except where physical limits were needed, like $\Omega_d>0$ and $\Omega_b>0$. {The chosen priors can be seen on Table \ref{tab:priors}.} In order to plot the results, we have used the free software getdist \cite{cosmomc}.}

\begin{table}[ht]
    \centering
    \begin{tabular} {| c|  c|}
    \hline
 Parameter &  Flat prior interval\\
\hline
{\boldmath$\Omega_{b0}       $} & $[0,0.2]$\\
{\boldmath$\Omega_{d0}       $} & $[0,1]$\\
{\boldmath$w       $} & $[-2,0]$\\
{\boldmath$\alpha         $} & $[-1,1]$\\
{\boldmath$H_0$} (km/s/Mpc) & $[20,120]$\\
\hline
\end{tabular}
    \caption{Chosen priors for the free parameters.}
    \label{tab:priors}
\end{table}

\begin{figure}[t!]
    \centering
    \includegraphics[width=\textwidth]{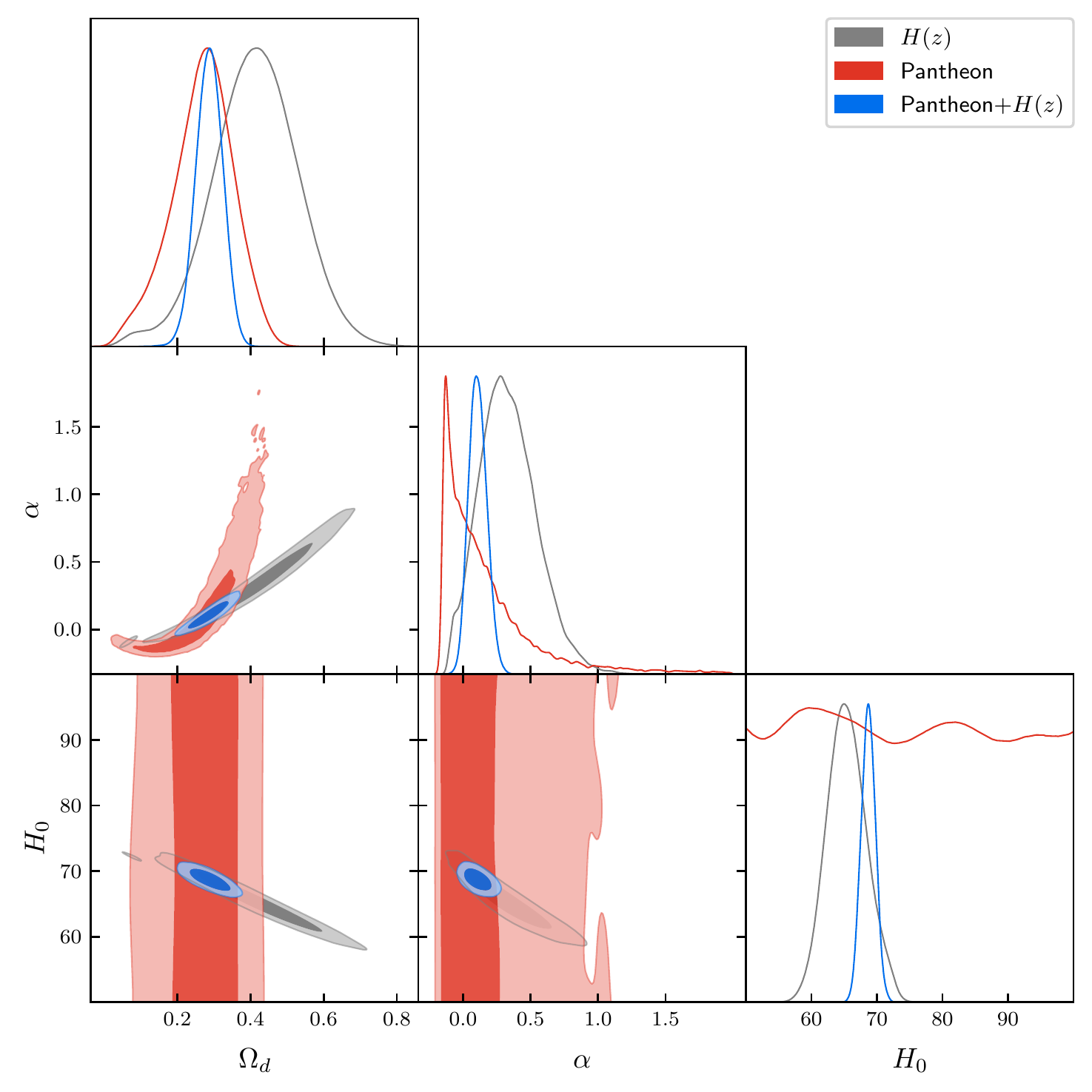}
    \caption{All constraints (SNe Ia and $H(z)$) {on} $\Lambda(t)$CDM model.}
    \label{fig:AllLtCDM}
\end{figure}

\begin{figure}[t!]
    \centering
    \includegraphics[width=\textwidth]{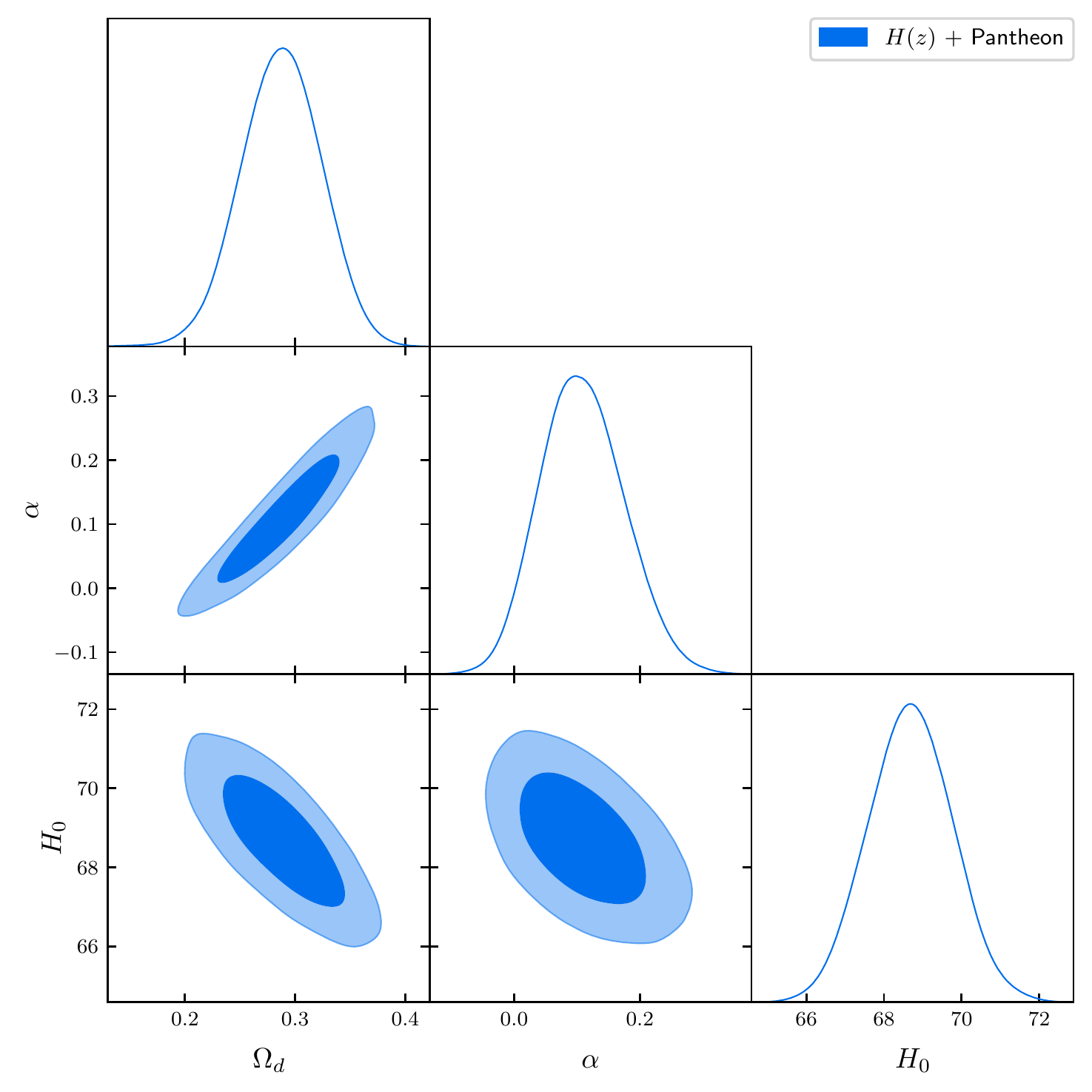}
    \caption{Combined constraints (SNe Ia and $H(z)$) {on} $\Lambda(t)$CDM model.}
    \label{fig:CombLtCDM}
\end{figure}

\subsection{\texorpdfstring{$\Lambda(t)$}{L(t)}CDM model}
In the case of $\Lambda(t)$CDM model, that is, interacting model where the dark energy has EOS $w=-1$, we have fixed the {baryon
density parameter} to {$\Omega_{b0}=0.0493$} according to CMB constraints from Planck (2018) \cite{Planck18}, which is also in agreement with Big Bang Nucleosynthesis (BBN) constraints. We then have as free parameters {$(H_0,\Omega_{d0},\alpha)$}.

The free parameters {are strongly} constrained from SNe Ia in combination to $H(z)$, as can be seen from Figs. \ref{fig:AllLtCDM} and \ref{fig:CombLtCDM} and Table \ref{tab:LtCDM}.

\begin{table}[ht]
    \centering
    \begin{tabular} {| c|  c|}
    \hline
 Parameter &  68\% and 95\% limits\\
\hline
{\boldmath$\Omega_{d0}       $} & $0.287^{+0.036+0.071}_{-0.036-0.075}   $\\
{\boldmath$\alpha         $} & $0.109^{+0.062+0.14}_{-0.072-0.12}      $\\
{\boldmath$H_0$} (km/s/Mpc) & $68.7^{+1.1+2.2}_{-1.1-2.1}        $\\
\hline
\end{tabular}
    \caption{Combined {constraints on} the free parameters of $\Lambda(t)$CDM model from Pantheon+$H(z)$ data for 68\% and 95\% C.L.}
    \label{tab:LtCDM}
\end{table}

As can be seen in Fig. \ref{fig:AllLtCDM}, mainly in the plane {$\Omega_{d0}$} -- $\alpha$, the combination {of} SNe Ia and $H(z)$ strongly reduces the parameter space. In Table \ref{tab:LtCDM}, the value for $H_0$ {is very close} with the value obtained for $\Lambda$CDM in Planck (2018) \cite{Planck18}, namely, $H_0=67.4\pm0.5$ km/s/Mpc. With our fixed value of {$\Omega_{b0}=0.0493$}, we obtain {$\Omega_{m0}=\Omega_{b0}+\Omega_{d0}=0.336\pm0.036$} at 68\% C.L., which {is very close} with the value obtained from Planck (2018) \cite{Planck18}, {$\Omega_{m0}=0.315\pm0.007$}. The value obtained for $\alpha$, as can be seen from Table \ref{tab:LtCDM}, is compatible with zero only with 1.6$\sigma$, so the interaction is not discarded by this analysis.

Fig. \ref{fig:CombLtCDM} shows in more detail the combined {constraints on} the free parameters {$(\Omega_{d0},\alpha,H_0)$}. We can see that the parameters are {very} constrained, with a larger correlation in the plane {$\Omega_{d0}$} -- $\alpha$.

{Ref. \cite{NunesValentino21} studies another $\Lambda(t)$CDM model, where they assume $Q=aH\xi\rho_x$ and $w=-0.999$. It is interesting to note that by using data from Pantheon, BAO and BBN, they find $\xi<0$, which implies $Q<0$, while we find $\alpha>0$, which implies $Q>0$ (DE decaying in DM). The difference between these results must come from the different interaction terms dependences.}

\subsection{Interacting XCDM model}
When we {leave} the dark energy EOS $w$ as a free parameter, allowing for $w\neq -1$,  {we have the so called XCDM model. In this case,} the SNe Ia+$H(z)$ data {are} not enough to constrain the free parameters {$(\Omega_{b0},\Omega_{d0},w,H_0)$} and we {have} to add constraints from other data. Thus, we have added constraints from Planck (2018) \cite{Planck18}, with the so called distance priors, which include the shift parameter, $\ell_A$ and $\Omega_{b0}h^2$. We have also included constraints from Baryon Acoustic Oscillations (BAO) from various surveys.

\begin{table}[ht]
    \centering
    \begin{tabular} {| c | c| c |}
    \hline
 Parameter & Flat $\Lambda$CDM prior & XCDM prior\\
\hline
{\boldmath$\Omega_{b0}       $} & $0.0496^{+0.0012+0.0024}_{-0.0012-0.0023}$ & $0.0497\pm0.0012\pm0.0023$\\
{\boldmath$\Omega_{d0}       $} & $0.2700^{+0.0062+0.013}_{-0.0062-0.012}$& $0.2696^{+0.0063+0.013}_{-0.0063-0.012}   $\\
{\boldmath$w         $} & $-1.071\pm0.045\pm0.090$& $-1.068^{+0.045+0.090}_{-0.045-0.091}  $\\
{\boldmath$\alpha$} & $-0.078^{+0.046+0.094}_{-0.046-0.092}  $& $-0.076\pm0.047\pm0.093$\\
{\boldmath$H_0$} (km/s/Mpc) & $67.13\pm0.74\pm1.5$& $67.09\pm0.74\pm1.5$\\
\hline
\end{tabular}
    \caption{Combined {constraints on} the free parameters of DM-DE interacting model from Pantheon+$H(z)$+Planck18+BAO.}
    \label{tab:XCDM}
\end{table}

\begin{figure}[t!]
    \centering
    \includegraphics[width=\textwidth]{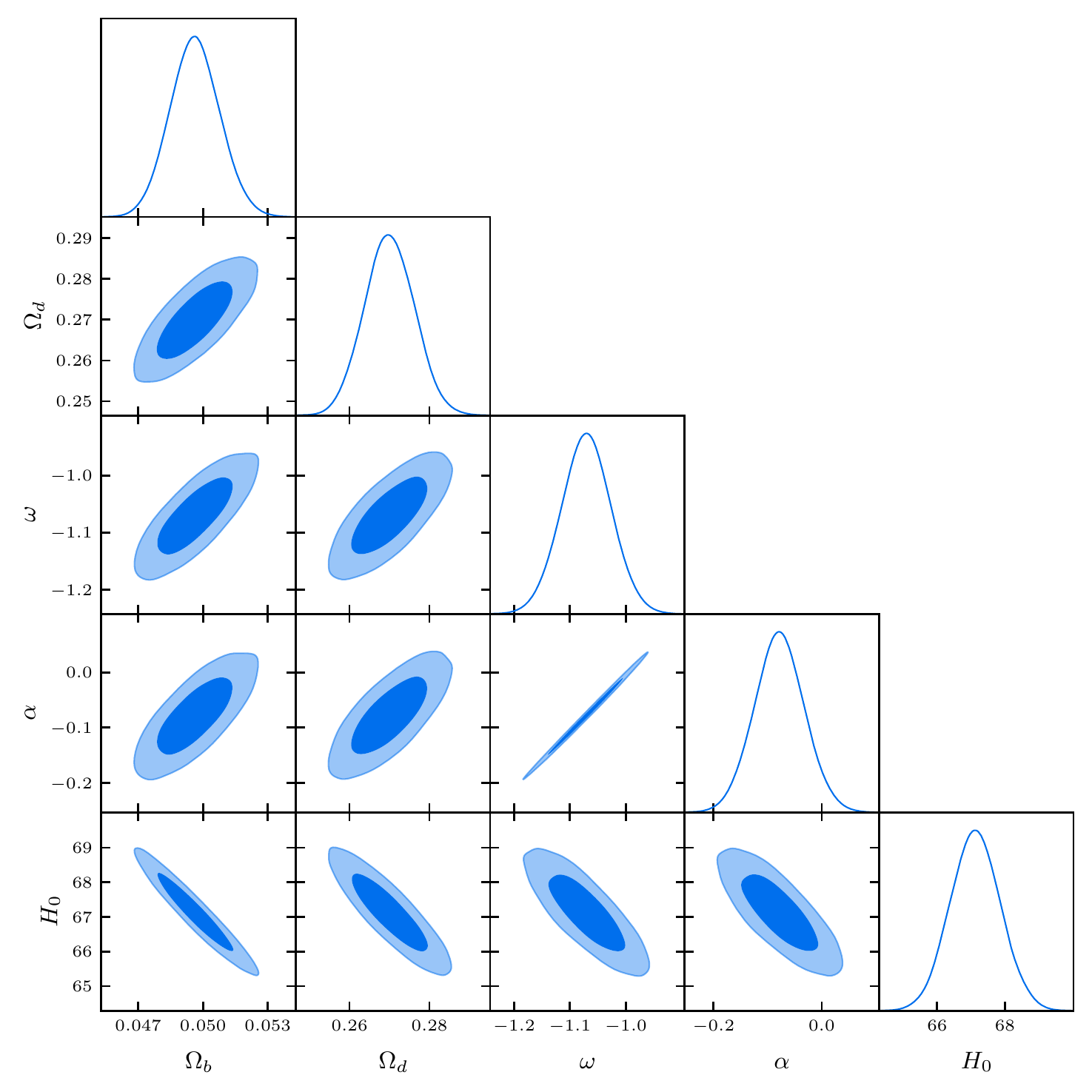}
    \caption{ {Combined constraints from Pantheon+$H(z)$+Planck18 ($\Lambda$CDM prior)+BAO {on} DM-DE interacting model {with $w\neq -1$}.}}
    \label{fig:DMDEComb}
\end{figure}

\begin{figure}[t!]
    \centering
    \includegraphics[width=\textwidth]{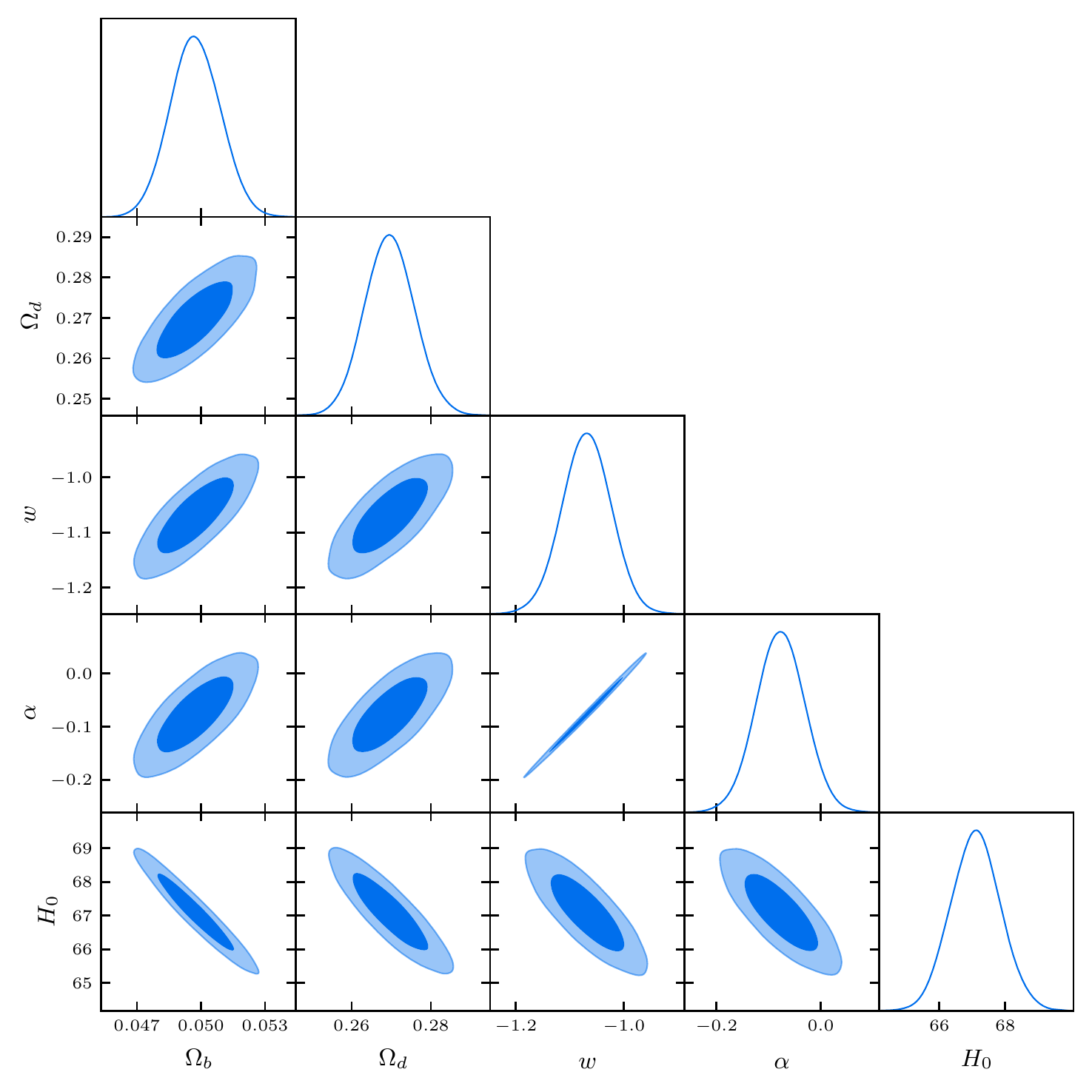}
    \caption{Combined constraints from Pantheon+$H(z)$+Planck18(XCDM prior)+BAO {on} DM-DE interacting model {with $w\neq -1$}.}
    \label{fig:DMDEComb2}
\end{figure}

{Table \ref{tab:XCDM} presents the combined  constraints  on  the  free  parameters  of  DM-DE  interacting  model  from Pantheon, $H(z)$, BAO and  Planck (2018) data for both prior models, {namely spatially flat $\Lambda$CDM and XCDM models}. From this Table, {Fig. \ref{fig:DMDEComb}} and {Fig. \ref{fig:DMDEComb2}}, the value obtained for {$\Omega_{m0}$} { is {$\Omega_{m0}=0.3196\pm0.0063$} for $\Lambda$CDM prior} and $\Omega_{m0}=0.3193\pm0.0064$ for XCDM prior, which are very close} to the value obtained from Planck (2018) \cite{Planck18}, {$\Omega_{m0}=0.3153\pm0.0073$}. {The values for $H_0$ are} also {compatible} with Planck (2018) \cite{Planck18}, $H_0=67.36\pm0.54$ km/s/Mpc, {for both priors}. {The results obtained for the DE EOS parameter, as can be seen in Table \ref{tab:XCDM}, is { compatible with $-1$ at 1.6$\sigma$ for $\Lambda$CDM prior} and  at 1.5$\sigma$ for XCDM prior. This shows that the phantom DE ($w<-1$) is slightly favoured by these analyses}. Other interesting feature of interacting DE is that the $\alpha$ parameter now changes sign when compared with $\Lambda(t)$CDM. { For the $\Lambda$CDM prior the value obtained is $\alpha=-0.078\pm0.046$, and for XCDM prior is $\alpha=-0.076\pm0.047$, which are marginally compatible with zero at 1.7 and 1.6$\sigma$, respectively.}

We also have made model comparisons using the Bayesian Information Criterion (BIC) \cite{Schwarz78,Liddle04,JesusEtAl16}, {among flat $\Lambda$CDM and $\Lambda(t)$CDM models. We have used, for this comparison, the Pantheon+$H(z)$ data. Similarly, it was also made a model comparison among Interacting XCDM and flat $\Lambda$CDM, using Pantheon+$H(z)$+Planck18(XCDM prior)+BAO data.} The results can be seen in Tables \ref{BICLtCDM} and \ref{BICXCDM}.

\begin{table}[h!]
\begin{tabular}{|l|c|c|c|c|c|c|c|}
\hline
Model    & Data                     & $\chi^2_{min}$ & $n_{par}$ & $n_{data}$ & BIC     & $\Delta$BIC & Support          \\ \hline
Flat $\Lambda$CDM  & Pantheon+$H(z)$            & 1057.74 & 2 & 1099  & 1071.74 & $0$    &  \\ 
$\Lambda(t)$CDM  & Pantheon+$H(z)$            & 1054.91 & 3    & 1099  & 1075.92 & $4.17$    & Strong to very strong/Significant  \\ \hline
\end{tabular}
\caption{BIC comparison for Pantheon+$H(z)$ data.}
\label{BICLtCDM}
\end{table}

\begin{table}[h!]
\begin{tabular}{|l|c|c|c|c|c|c|c|}
\hline
Model    & Data                     & $\chi^2_{min}$ & $n_{par}$ & $n_{data}$ & BIC     & $\Delta$BIC & Support          \\ \hline
Flat $\Lambda$CDM  & PHPB\footnote{Pantheon+$H(z)$+Planck+BAO} & 1064.84 & 3 & 1079  & 1085.79 & $0$    & \\
XCDM+Int & PHPB & 1059.28 & 5    & 1079  & 1094.20 & 8.41 & Decisive/Strong  \\ \hline

\end{tabular}
\caption{{ BIC comparison for Pantheon+$H(z)$+Planck18(XCDM prior)+BAO data.}}
\label{BICXCDM}
\end{table}

As one can see from Table \ref{BICXCDM}, the Interacting XCDM model can be discarded in comparison to {flat $\Lambda$CDM model by this analysis.} In the case of $\Lambda(t)$CDM (Table \ref{BICLtCDM}), however, {the model is significantly disfavoured when compared to flat $\Lambda$CDM.} That is, this analysis shows that Pantheon+$H(z)$ data prefer this interacting $\Lambda$CDM model.

\section{Conclusion}
Was CP solved or alleviated? {In order to try to answer this question, we have studied a model of interacting DE-DM where the ratio between these components can change from early to late times, satisfying a relation of the form $\rho_x\propto\rho_d^\alpha$, where $\rho_x$ and $\rho_d$ are the energy densities of dark energy and dark matter components, respectively, and $\alpha$ is a free parameter. Then, we compared the ratio $\rho_d/\rho_x$ along time for two distinct models, namely the $\Lambda(t)$CDM model and interacting XCDM model. We say that CP is solved if the ratio is $\sim 1$ along time evolution, and it is alleviated if the ratio is less than that predicted by the current standard model.}

In the case of $\Lambda(t)$CDM, at the 2$\sigma$ {interval} of parameter $\alpha$, the ratio $\rho_d/\rho_\Lambda$ at $z=1000$ is reduced from $\approx10^9$ to $\approx3\times10^5$, and CP is alleviated. For interacting XCDM, however, a degeneracy emerges on the plane $w-\alpha$, in such a way that for phantom DE, $\alpha$ becomes negative. For the best fit, the CP is even worse, $\rho_d/\rho_x\approx2\times10^9$ at $z=1000$. At the 2$\sigma$ {interval} of parameter $\alpha$, the ratio $\rho_d/\rho_x$ is slightly reduced to $\approx2\times10^8$, and CP is not even alleviated. Yet, only if one would make a full CMB spectrum analysis, a final conclusion could be drawn, in this case.

{An interesting argument about interaction terms comes from \cite{Valiviita2008}, where they argue that $Q$ should not involve a global quantity like $H$, as the DE-DM interaction should occur locally. As defined above from Eqs. \eqref{Qrhox} and \eqref{Gamma}, our interaction term accomodates both possibilities: If $w\neq-1$, $Q$ depends on $H$. For $\Lambda(t)$CDM, however, it does not and can be interpreted as local.}

Another interesting feature is that, as the model with $w \neq -1$ extends to values with phantom energy solutions ($w \leq -1$), there is the possibility of the decay of dark energy in dark matter restrain the Big Rip singularity, either by changing the whole evolution of the expansion or causing great production of dark matter at the last moments, according to the choice of parameters. Similarly, there is a possibility of having a Big Rip even with $w \geq -1$, if dark matter is decaying into dark energy at a rate large enough. Such possibilities can be analyzed in future works.

\appendix
\section{Analytical solution for \texorpdfstring{$w=-1$}{w=-1}}

Eq. (\ref{oderhodw}) with $w=-1$ is:

\begin{equation}
     (1+\alpha\beta{\rho_{d}}^{\alpha-1})\frac{d\rho_{d}}{dt}=-\frac{3\rho_{d}}{a}\frac{da}{dt}\,,
      \label{12}
\end{equation}
which can be written as:
\begin{equation}
\left(\frac{1}{\rho_{d}}+\alpha\beta{\rho_{d}}^{\alpha-2}\right)d\rho_{d}=-\frac{3da}{a}\,.
    \label{13}
\end{equation}
The solution of this equation is:
\begin{equation}
    \ln{\frac{\rho_{d}}{\rho_{d0}}}+\frac{\alpha\beta}{\alpha-1}(\rho_{d}^{\alpha-1}-\rho_{d0}^{\alpha-1})=-3\ln{a},
    \label{17}
\end{equation}
where we have used the initial condition $\rho_d(a=1)=\rho_{d0}$. It can also be written as:
\begin{equation}
    \ln{\rho_d^{\alpha-1}}+\alpha\beta\rho_d^{\alpha-1}=(\alpha-1)\ln{(\rho_{d0}a^{-3})}+\alpha\beta\rho_{d0}^{\alpha-1}
    \label{23}
\end{equation}

In order to solve \eqref{23}, we now introduce the Lambert function $W$ {\cite{CorlessEtAl96}} defined as a solution of the equation:
\begin{equation}
    xe^{x}=f\Rightarrow x=W(f)
    \label{20}
\end{equation}
or
\begin{equation}
    x+\ln{x}=\ln{f} .
    \label{21}
\end{equation}

As shown by {\cite{CorlessEtAl96,Lemeray1897}}, besides being used to solve Eqs. \eqref{20}-\eqref{21}, the Lambert $W$ function can also be used to solve
\be
xB^x=A,
\label{powx}
\ee
with solution
\be
x=\frac{W(A\ln B)}{\ln B}
\ee
Eq. \eqref{powx} can also be written as:
\be
\ln x+x\ln B=\ln A
\label{xln}
\ee

Comparing \eqref{xln} with \eqref{23}, we may identify $x=\rho_d^\alpha$, $\ln B=\alpha\beta$ and $\ln A=(\alpha-1)\ln{(\rho_{d0}a^{-3})}+\alpha\beta\rho_{d0}^{\alpha-1}$, so that we find the solution:
\be
\rho_d=\left[\frac{W\left(\alpha\beta\exp(\alpha\beta\rho_{d0}^{\alpha-1}) (\rho_{d0}a^{-3})^{\alpha-1}\right)}{\alpha\beta}\right]^{\frac{1}{\alpha-1}}
\label{eqsol}
\ee

Defining the ratio $r\equiv\frac{\rho_x}{\rho_d}$, the relation \eqref{rhoxd}, yields $r=\beta\rho_d^{\alpha-1}$, so that we can use the DE-DM ratio today, $r_0=\beta\rho_{d0}^{\alpha-1}$, to rewrite the solution \eqref{eqsol} as
\be
\rho_x=\rho_{x0}\left[\frac{W\left(\alpha r_0 e^{\alpha r_0} a^{3(1-\alpha)}\right)}{\alpha r_0}\right]^{\frac{\alpha}{\alpha-1}}\,,
\ee

\begin{acknowledgements}
This study was financed in part by the Coordena\c{c}\~ao de Aperfei\c{c}oamento de Pessoal de N\'ivel Superior - Brasil (CAPES) - Finance Code 001. AAE and DB would like to thank CAPES. SHP would like to thank CNPq - Conselho Nacional de Desenvolvimento Cient\'ifico e Tecnol\'ogico, Brazilian research agency, for financial support, grants number 303583/2018-5 and 308469/2021-6.
\end{acknowledgements}

\end{document}